\documentclass[pmlr]{jmlr}

\usepackage{longtable}

\usepackage{booktabs}
\usepackage[load-configurations=version-1]{siunitx} 
\usepackage{times}
\usepackage{latexsym}
\usepackage{graphicx}
\usepackage{listings}
\usepackage{amsmath}
\usepackage[T1]{fontenc}
\usepackage{pifont}
\usepackage{tcolorbox}
\usepackage{float}
\usepackage[utf8]{inputenc}

\usepackage{microtype}

\usepackage{inconsolata}

\theorembodyfont{\upshape}
\theoremheaderfont{\scshape}
\theorempostheader{:}
\theoremsep{\newline}

\newtcolorbox{boxK}{
    sharpish corners, 
    boxrule = 0pt,
    toprule = 4.5pt, 
    enhanced,
    fuzzy shadow = {0pt}{-2pt}{-0.5pt}{0.5pt}{black!35} 
}

\jmlrvolume{1}
\jmlryear{2024}
\jmlrworkshop{AI for Education}

\title[Automated Assessment of Students' Code Comprehension using LLMs]{Automated Assessment of Students' Code Comprehension using LLMs}

 \author{
  \Name{Priti Oli} \Email{poli@memphis.edu}\\ 
  \Name{Rabin Banjade} \Email{rbnjade1@memphis.edu}\\
  \Name{Jeevan Chapagain} \Email{jchpgain@memphis.edu}\\
  \Name{Vasile Rus} \Email{vrus@memphis.edu}\\
  \addr University of Memphis, Memphis TN 38152,USA
  }

\begin{document}

\maketitle

\begin{abstract}

 Assessing student's answers and in particular natural language answers is a crucial challenge in the field of education. Advances in machine learning, including transformer-based models such as Large Language Models(LLMs), have led to significant progress in various natural language tasks. Nevertheless, amidst the growing trend of evaluating LLMs across diverse tasks, evaluating LLMs in the realm of automated answer assesment has not received much attention. To address this gap, we explore the potential of using LLMs for automated assessment of student's short and open-ended answer. Particularly, we use LLMs to compare students' explanations with expert explanations in the context of line-by-line explanations of computer programs.
 For comparison purposes, we assess both Large Language Models (LLMs) and encoder-based Semantic Textual Similarity (STS) models in the context of assessing the correctness of students' explanation of computer code. 
Our findings indicate that LLMs, when prompted in few-shot and chain-of-thought setting perform comparable to fine-tuned encoder-based models in evaluating students' short answers in programming domain.


\end{abstract}

\section{Introduction}
\label{sec:intro}

Large Language Models (LLMs), such as ChatGPT, have garnered significant attention for their remarkable ability to generate responses to user prompts. These models have been explored for their potential (and risks) for education, particularly in the realm of computer science (CS) education, which is our focus. For CS education, LLMs have been investigated with respect to creating programming exercises~\citep{sarsa22}, generating code explanations~\citep{mcneil23}, and even providing assistance in debugging coding problems~\citep{liffiton2023codehelp}, among other educational applications. While numerous studies have highlighted ChatGPT's generative capabilities of educational resources and assistance, there is a notable gap in exploring ChatGPT's assessment capabilities within educational contexts. In this work, we evaluate the effectiveness of LLMs to automatically assess students' self-explanations of code. Such explanations are generated, for instance, while students engage in code comprehension activities with a computer tutor, which needs to automatically assess the correctness of students' explanations of code in order to provide feedback. It should be noted that self-explanation, i.e., explaining learning material to oneself through speaking or writing ~\citep{mcnamara2009self}, has been shown to improve comprehension and learning of programming concepts in introductory computer science courses ~\citep{tamang2021comparative,oli2023improving}. 


A simple and scalable approach to assessing student explanations is semantic similarity, i.e. measuring the similarity of such natural language explanations to an appropriate reference/correct answer, e.g., provided by an expert, through an automated short answer grading system~\citep{mohler2009text}. If the student's self-explanation is semantically similar to the reference answer the student's self-explanation is deemed to be correct. 


Semantic similarity measures the degree to which two fragments of text have similar meanings by producing a similarity score, ranging from 0 to 1 (normalized score), 0 meaning no similarity at all, whereas 1 meaning semantically equivalent~\citep{cer2017semeval}. Although there have been numerous studies~\citep{cer2017semeval} measuring semantic similarity between texts, limited research has been conducted in the area of computer programming and source code comprehension. In our study, we employ Large Language Models (LLM) to automatically evaluate students' line-by-line explanation of code and compare it with encoder-based models. We evaluate the proposed LLM-based approach using a set of student self-explanations produced in the context of an online learning environment that asks students to freely explain Java code examples line-by-line. This is part of a broader project to develop an educational technology that can monitor, model, and scaffold students' code comprehension processes.

\section{Related Work}
\label{sec:related_work}
\

\textbf{Automated Short Answer Grading (ASAG).}
Prior work on ASAG has been based on determining the semantic similarity between learner answers and reference answers in various domains such as Physics, Biology, Geometry etc.~\citep{leacock2003c, mohler2009text}.
Early methods for measuring semantic similarity for ASAG relied on using hand-crafted features, such as syntactic feature~\citep{leacock2003c}, lexical similarity~\citep{dzikovska2012towards}, n-gram features~\citep{heilman2013ets}, vector-based similarity features~\citep{sultan2016fast} and graph alignment feature~\citep{mohler2011learning} to automatically assess student's short answer.
The advances in neural networks led to the introduction of numerous deep learning-based Automated Short Answer Grading systems~\citep{dasgupta2018augmenting,pontes2018predicting} .

Previous studies investigating pre-trained transformers in NLP tasks have observed significant performance improvements in Automated Short Answer Grading ~\citep{camus2020investigating} through fine-tuning on datasets such as MNLI~\citep{williams2017broad} and SemEval-2013~\citep{segura2013semeval}.
\cite{lun2020multiple} used fine-tuned BERT model along with multiple data augmentation technique for automatic short answer scoring. 
~\cite{khayi2021towards} proposed fine-tuning pre-trained transformers for the automated evaluation of freely generated student responses in Physics, implemented within a dialogue-based Intelligent Tutoring System. 
~\citep{sung2019pre} fine-tuned BERT on domain specific data such as textbooks and reported that fine-tuning a pre-trained model for task-specific purpose demonstrates superior performance in short answer grading. Along those lines, in our study, we fine-tune pre-trained models for assessing short answer in program comprehension, an area which has not been previously investigated. Furthermore, we propose a novel method of employing LLM towards short answer assessment and compare it's performance with the encoder-based transformer model. It should be noted that in the programming-comprehension domain, ~\cite{fowler2021autograding} trained a bag-of-word and bigram model to automatically assess students' explanation of code, i.e., their method does not rely on latest, most promising, deep learning methods for short natural language student answers.


\textbf{Evaluating LLMs on Semantic Similarity Task.} 
In their study, ~\cite{zhong2023can} report that ChatGPT surpasses all BERT models with a substantial margin in an inference task and attains comparable performance to BERT in sentiment analysis and question-answering tasks. However, their study indicates that ChatGPT has limited ability in paraphrase and semantic similarity tasks. However, ~\cite{gatto2023text} demonstrate that the Semantic Textual Similarity (STS) task can be effectively framed as a text generation problem, achieving robust performance with LLM outperforming encoder-based STS models across various STS benchmarks.

Given that LLMs benefit significantly from training on code and its corresponding summaries, in this study we investigate the applicability of LLMs to automatically assess students' line-by-line explanations of code.

\vspace{-4mm}
\section{Dataset}
\label{sec:dataset}
\vspace{-3mm}
\begin{table*}[h!]
  \centering
  \begin{tabular}{p{4cm}|p{4cm}|p{3.5cm}|c}
    Code Snippet & Expert Explanation & Student Explanation & Similarity\\

  \hline


lst[i] += 1;
  
  & Increment the current element in the array by 1.
&  This statement increments the element at the index i of the array by 1.  & 5
\\

int minutes = seconds / 60;
& To obtain the minutes in seconds, we divide the minutes by 60 because there are 60 seconds in a minute  & Create the variable minutes & 2

  \end{tabular}
  \caption{An example of the code, student's explanation, expert's explanation and Similarity Score }
  \label{tab:dataset_example}
\end{table*}

In our work, we use the \textit{CodeCorpus}\footnote{anonymized for the blind review process} for our analysis. CodeCorpus was developed with the goal of understanding how learners explain code. It consists of pairs of code snippets accompanied by expert explanations and explanations given by students/non-experts for ten different code examples. To collect the student's explanations, Amazon Mechanical Turk (MTurk) was used. The MTurk HIT (Human Intelligence Task) to collect learners' self-explanations was available only to workers from the United States and Canada who had to qualify for the task by correctly answering 2 out of 3 multiple choice basic program construction tasks that involved selecting the correct missing line. The qualification test was needed to make sure participants have some minimal programming background. 
Expert explanations were acquired from a curated collection of annotated examples within a comprehensive repository of interactive learning content~\citep{hicks2020live}. 
These expert explanations serve as reference explanations when assessing learners' self-explanations.

In addition to expert explanations, human judgments of the semantic similarity between the expert and stundet's code explanations were obtained. Three graduate students in Computer Science annotated on a scale of 1-5 about 1,770 pairs of expert and student's explanations which are used in our study presented here. In the data set, 18\% of the sentence pairs scored 4 or 5 (high semantic similarity), while 59\% were labeled incorrect (score 1) or exhibited low concept coverage (score 2). About 23\% of the sentence pairs received a score of 3. The semantic similarity label distribution is shown in Table~\ref{tab:data_distribution} and Table ~\ref{tab:dataset_example} provides example instances from the corpus. Given the opaque nature of ChatGPT's training data, we validate our findings against memorization by exclusively working with publicly released dataset after May 2023.  

\section{Methodology}
 \label{sec:methodology}


As already noted, we employ semantic similarity to evaluate students' natural language responses, with the primary focus on assessing LLMs' capability in measuring semantic similarity; however, for comparison purposes, we offer results with several other approaches, as described next.

\vspace{-2mm}
\subsection{Assessment Using Encoder Models}


First, we calculate the similarity based on BERTScore ~\citep{zhang2019bertscore} and Universal Sentence Encoder (USE; ~\citep{cer2018universal}). Second, we employ SentenceBERT(SBERT) models ~\citep{reimers2019sentence}, which we fine-tune on our dataset. The three fine-tuned models include: i) RoBERTa fine-tuned on NLI, ii) CodeBERT, and iii) all-mpnet-base-v2\footnote{https://huggingface.co/sentence-transformers/all-mpnet-base-v2}. We experimented with CodeBERT~\citep{feng2020codebert} as an encoder to assess whether it offers advantages in capturing the similarity of sentences related to code segments. Also, since SBERT models show performance improvement if we fine-tune it in previously fine-tuned NLI data~\citep{reimers-2019-sentence-bert} fine tuned models, we employed models fine-tuned in NLI to further fine-tune using our dataset. For each of the mentioned encoders, we compute the similarity between expert and student explanations by calculating the cosine similarity between their embeddings. There is an exception to this similarity computation when calculating BERTScore. In this case, the similarity of two sentences is computed as the sum of cosine similarities between their tokens' embeddings.

We finetune SBERT with contrastive loss objective function for one epoch. We used a batch size of 16, Adam optimizer with a learning rate $2 {e^{-4}}$ and a linear learning rate warm-up over 10\% of the training data. Our pooling strategy is MEAN. This comprehensive assessment framework allows us to thoroughly evaluate the effectiveness of different language models and baselines in capturing semantic similarity in the context of answer assessment.

\subsection{Assessment by Prompting LLMs:}
We explore various prompting strategies for four different large language models: OpenAI's ChatGPT-3.5-turbo-0613 and ChatGPT-4-0613 ~\citep{openai2023gpt4}, gpt-4-1106-preview(GPT-4 Turbo)~\citep{openai2023gpt4} and Meta's open source model LLMa2-chat\footnote{https://huggingface.co/meta-llama/Llama-2-7b-chat-hf}~\citep{llama2}).

First, for predictive prompting of semantic similarity, we used simple prompts to instruct the LLM to predict the similarity score on a scale of 1-5, similar to human judgments (with 1 indicating no semantic similarity and 5 indicating semantically equivalence between the pair of sentences). Based on the findings by ~\cite{gatto2023text}, who suggest framing STS tasks to predict a similarity percentage (leveraging large language models' strong textual reasoning and their exposure to percentage-related language during pre-training), we further used the same prompt to generate the semantic similarity in the scale of 0-1.
In addition, we also explore advanced prompting strategies. These include the conventional few-shot prompting, also known as in-context learning, where the LLM is tasked to infer from the provided examples or task descriptions ~\citep{brown2020language}, as well as  few-shot chain-of-thought (CoT) prompting~\citep{wei2022chain} where the LLM is guided to think step by step. In the case of few-shot learning, we employed a stratified sampling approach to select six expert explanations along with corresponding student explanations and benchmark similarity scores. These were provided as examples to the Large Language Models (LLMs), with the caveat that the examples were excluded from the dataset used for subsequent analysis.

For few-shot Chain-of-Thought prompting, we manually crafted a step-by-step breakdown of the reasoning behind assigning semantic similarity scores when evaluating two texts, selecting three examples with varying benchmark similarity scores. The prompts utilized in our analysis are detailed in Appendix \ref{apd:prompts}. In the CoT Prompting approach, which elicited textual responses along with reasoning, we extracted numerical values within specified delimiters to obtain the semantic similarity score.
In our experimental setup, we opted for deterministic results by setting the temperature parameter to 0. We set a maximum token length of 1200 to limit the scope of generated sequences.


\label{sec:citep}
\section{Results and Discussion}

\begin{table*}[t]
   \centering
    \begin{tabular}{cc|c|c}
          & \textbf{Model}  &\textbf{Pearson} & \textbf{Spearman}  \\
         \hline

          & BERTScore  &0.573 & 0.553  \\
          & USE  & 0.61 & 0.61 \\

\midrule
          Sentence Transformer & RoBERTa-base\textsuperscript{{$\dagger$}} &  0.800   &    0.78 \\
         &CodeBERT-base\textsuperscript{{$\dagger$}}  & 0.797 & 0.761  \\
          &all-mpnet\textsuperscript{{$\dagger$}}  &  \textbf{0.81} & \textbf{0.811}  \\

\midrule

       GPT-3.5 &baseline-prompt[1-5]  & 0.58 & 0.59 \\
         &baseline-prompt[0-1]  & 0.60 & 0.61 \\
         &fewshot-prompt[0-1]  & 0.64 & 0.64 \\
         &CoT-prompt[0-1]  & 0.69 & 0.70\\

\midrule
         GPT-4 & baseline-prompt[1-5] & 0.69 & 0.70 \\
         &baseline-prompt[0-1]  & 0.72 & 0.737 \\
         &fewshot-prompt[0-1]  & 0.78 & 0.79 \\
         &CoT-prompt[0-1]  & \textbf{0.81} & \textbf{0.82} \\

\midrule

       GPT-4-turbo &baseline-prompt[1-5]  & 0.70 & 0.70 \\
         &baseline-prompt[0-1] & 0.72 & 0.75 \\
         &fewshot-prompt[0-1]  & 0.67 & 0.71 \\
         &CoT-prompt[0-1]  &  0.79 & 0.80 \\

\midrule
        LLAMA-2 & baseline-prompt[1-5]  & 0.29 & 0.31 \\
        & baseline-prompt[0-1]  &  0.38 & 0.39 \\
        & few-shot-prompt[0-1]  & 0.42 & 0.44 \\
        & CoT-prompt[0-1]  & 0.26 & 0.27 \\

    \end{tabular}
        \caption{Pearson and Spearman correlations by comparing human-annotated semantic similarity scores with automated similarity scores for student and expert explanations across different model classes. {$\dagger$} indicate finetuned model }
        \label{tab:results}
        \vspace{-5mm}
\end{table*}

\subsection{Assessment using Encoder-based Model}

As we can see from results in Table \ref{tab:results}, for encoder based models, models such as BERTScore and Universal Sentence Encoder show below par results based on Pearson and Spearman rank correlation. The results indicate that fine-tuned sentence transformer models capture the assessment score better. The encoder models pre-trained on code such as CodeBERT do not show better performance compared to RoBERTa. The best performing model for student answer explanation is \textit{all-mpnet}. One of the reasons for this might be the large amount of data it is fine tuned on. Also, there is no remarkable difference between RoBERTa and \textit{all-mpnet} indicating sentence transformer models can be used effectively for student answer assessment by comparing expert explanations with student explanations.

\subsection{Assessment by Prompting LLM}

In Table~\ref{tab:results}, the results of prompting Large Language Models (LLM) to assess semantic similarity are presented. The outcomes for various versions of ChatGPT indicate a notable trend: prompting the LLMs to predict semantic similarity on a scale of 0-1 yields superior performance compared to prompting it to predict similarity on other arbitrary scales(1-5).

The advance strategies consistently boost ChatGPT's performance, with manual chain-of-thought (CoT) providing the most significant benefits. Notably, the standard few-shot CoT enhances ChatGPT's overall performance (on average 15\% better than baseline prompting for ChatGPT based model) with GPT-4  providing the best performance for our task. Table~\ref{tab:results} shows that GPT4 performs similarly to fine-tuned encoder-based models when using chain-of-thought prompting. The results also indicate that GPT-4 consistently outperforms GPT-3.5 across various prompting techniques and scales. Our experimentation with GPT-4 Turbo yielded results comparable to those of other LLMs, offering no discernible advantage except processing speed. ChatGPT-4 demonstrates superior reasoning in CoT-prompting and also closely aligns with human-annotated benchmark similarity (see Appendix \ref{apd:cot_response} for examples). 
Concerning LLama-2, the semantic similarity scores were skewed towards higher values, particularly with scores of 0.8-1.0 in the scale of 0-1. Moreover, when prompted to generate only the semantic similarity score, LLama-2 produced excessive text, including unnecessary reasoning about the semantic similarity score.


\subsection{Error Analysis}
When prompting LLMs, we found in-context learning to be sensitive to the provided examples, which is consistent with the findings from previous studies~\citep{agrawal2022context,zhong2023can}. This sensitivity may arise from limited generalization or overfitting to few-shot examples used, suggesting a potential correlation between provided examples and test data. To address this potential bias in the few-shot setting, we conducted the analysis three times with different instances of example provided each time  and present the results as the average of these runs.

One of the cases where the LLMs fail is for instances involving numerical reasoning. LLMs assign a high semantic equivalence score to such instances, which although they are linguistically highly similar, often involve different numerical values. For example, the LLMs scored the student's explanation of \textit{``creates variable integer entitled `num' with initial value 5"} with a similarity of 0.8 compared to the reference \textit{``In this program, we initialize the variable num to 15." }(for more detail see Appendix ~\ref{apd:error_analysis}). In this scenario, the student's response suggests a potential gap in understanding, highlighting the need for an instructional intervention. 

\section{Conclusion }
This work investigated the ability of LLMs to automatically assess students' code comprehension.
Our results indicate that Large Language Models (LLMs) perform comparably well, in particular GPT models, to fine-tuned encoder-based models but there is room for improvement which we plan to explore in the future. 


\bibliography{bibliography}

\begin{thebibliography}{35}
\providecommand{\natexlab}[1]{#1}
\providecommand{\url}[1]{\texttt{#1}}
\expandafter\ifx\csname urlstyle\endcsname\relax
  \providecommand{\doi}[1]{doi: #1}\else
  \providecommand{\doi}{doi: \begingroup \urlstyle{rm}\Url}\fi

\bibitem[Agrawal et~al.(2022)Agrawal, Zhou, Lewis, Zettlemoyer, and Ghazvininejad]{agrawal2022context}
Sweta Agrawal, Chunting Zhou, Mike Lewis, Luke Zettlemoyer, and Marjan Ghazvininejad.
\newblock In-context examples selection for machine translation.
\newblock \emph{arXiv preprint arXiv:2212.02437}, 2022.

\bibitem[Brown et~al.(2020)Brown, Mann, Ryder, Subbiah, Kaplan, Dhariwal, Neelakantan, Shyam, Sastry, Askell, et~al.]{brown2020language}
Tom Brown, Benjamin Mann, Nick Ryder, Melanie Subbiah, Jared~D Kaplan, Prafulla Dhariwal, Arvind Neelakantan, Pranav Shyam, Girish Sastry, Amanda Askell, et~al.
\newblock Language models are few-shot learners.
\newblock \emph{Advances in neural information processing systems}, 33:\penalty0 1877--1901, 2020.

\bibitem[Camus and Filighera(2020)]{camus2020investigating}
Leon Camus and Anna Filighera.
\newblock Investigating transformers for automatic short answer grading.
\newblock In \emph{Artificial Intelligence in Education: 21st International Conference, AIED 2020, Ifrane, Morocco, July 6--10, 2020, Proceedings, Part II 21}, pages 43--48. Springer, 2020.

\bibitem[Cer et~al.(2017)Cer, Diab, Agirre, Lopez-Gazpio, and Specia]{cer2017semeval}
Daniel Cer, Mona Diab, Eneko Agirre, Inigo Lopez-Gazpio, and Lucia Specia.
\newblock Semeval-2017 task 1: Semantic textual similarity-multilingual and cross-lingual focused evaluation.
\newblock \emph{arXiv preprint arXiv:1708.00055}, 2017.

\bibitem[Cer et~al.(2018)Cer, Yang, Kong, Hua, Limtiaco, John, Constant, Guajardo-Cespedes, Yuan, Tar, et~al.]{cer2018universal}
Daniel Cer, Yinfei Yang, Sheng-yi Kong, Nan Hua, Nicole Limtiaco, Rhomni~St John, Noah Constant, Mario Guajardo-Cespedes, Steve Yuan, Chris Tar, et~al.
\newblock Universal sentence encoder.
\newblock \emph{arXiv preprint arXiv:1803.11175}, 2018.

\bibitem[Dasgupta et~al.(2018)Dasgupta, Naskar, Dey, and Saha]{dasgupta2018augmenting}
Tirthankar Dasgupta, Abir Naskar, Lipika Dey, and Rupsa Saha.
\newblock Augmenting textual qualitative features in deep convolution recurrent neural network for automatic essay scoring.
\newblock In \emph{Proceedings of the 5th Workshop on Natural Language Processing Techniques for Educational Applications}, pages 93--102, 2018.

\bibitem[Dzikovska et~al.(2012)Dzikovska, Nielsen, and Brew]{dzikovska2012towards}
Myroslava~O Dzikovska, Rodney~D Nielsen, and Chris Brew.
\newblock Towards effective tutorial feedback for explanation questions: A dataset and baselines.
\newblock In \emph{Proceedings of the 2012 conference of the North American Chapter of the Association for Computational Linguistics: Human language technologies}, pages 200--210. Association for Computational Linguistics, 2012.

\bibitem[Feng et~al.(2020)Feng, Guo, Tang, Duan, Feng, Gong, Shou, Qin, Liu, Jiang, et~al.]{feng2020codebert}
Zhangyin Feng, Daya Guo, Duyu Tang, Nan Duan, Xiaocheng Feng, Ming Gong, Linjun Shou, Bing Qin, Ting Liu, Daxin Jiang, et~al.
\newblock Codebert: A pre-trained model for programming and natural languages.
\newblock \emph{arXiv preprint arXiv:2002.08155}, 2020.

\bibitem[Fowler et~al.(2021)Fowler, Chen, Azad, West, and Zilles]{fowler2021autograding}
Max Fowler, Binglin Chen, Sushmita Azad, Matthew West, and Craig Zilles.
\newblock Autograding" explain in plain english" questions using nlp.
\newblock In \emph{Proceedings of the 52nd ACM Technical Symposium on Computer Science Education}, pages 1163--1169, 2021.

\bibitem[Gatto et~al.(2023)Gatto, Sharif, Seegmiller, Bohlman, and Preum]{gatto2023text}
Joseph Gatto, Omar Sharif, Parker Seegmiller, Philip Bohlman, and Sarah~Masud Preum.
\newblock Text encoders lack knowledge: Leveraging generative llms for domain-specific semantic textual similarity.
\newblock \emph{arXiv preprint arXiv:2309.06541}, 2023.

\bibitem[Heilman and Madnani(2013)]{heilman2013ets}
Michael Heilman and Nitin Madnani.
\newblock Ets: Domain adaptation and stacking for short answer scoring.
\newblock In \emph{Second Joint Conference on Lexical and Computational Semantics (* SEM), Volume 2: Proceedings of the Seventh International Workshop on Semantic Evaluation (SemEval 2013)}, pages 275--279, 2013.

\bibitem[Hicks et~al.(2020)Hicks, Akhuseyinoglu, Shaffer, and Brusilovsky]{hicks2020live}
Alexander Hicks, Kamil Akhuseyinoglu, Clifford Shaffer, and Peter Brusilovsky.
\newblock Live catalog of smart learning objects for computer science education.
\newblock In \emph{Sixth SPLICE Workshop}, 2020.

\bibitem[Khayi et~al.(2021)Khayi, Rus, and Tamang]{khayi2021towards}
Nisrine~Ait Khayi, Vasile Rus, and Lasang Tamang.
\newblock Towards improving open student answer assessment using pretrained transformers.
\newblock In \emph{The International FLAIRS Conference Proceedings}, volume~34, 2021.

\bibitem[Leacock and Chodorow(2003)]{leacock2003c}
Claudia Leacock and Martin Chodorow.
\newblock C-rater: Automated scoring of short-answer questions.
\newblock \emph{Computers and the Humanities}, 37:\penalty0 389--405, 2003.

\bibitem[Liffiton et~al.(2023)Liffiton, Sheese, Savelka, and Denny]{liffiton2023codehelp}
Mark Liffiton, Brad Sheese, Jaromir Savelka, and Paul Denny.
\newblock Codehelp: Using large language models with guardrails for scalable support in programming classes.
\newblock \emph{arXiv e-prints}, pages arXiv--2308, 2023.

\bibitem[Lun et~al.(2020)Lun, Zhu, Tang, and Yang]{lun2020multiple}
Jiaqi Lun, Jia Zhu, Yong Tang, and Min Yang.
\newblock Multiple data augmentation strategies for improving performance on automatic short answer scoring.
\newblock In \emph{Proceedings of the AAAI Conference on Artificial Intelligence}, volume~34, pages 13389--13396, 2020.

\bibitem[McNamara and Magliano(2009)]{mcnamara2009self}
Danielle~S McNamara and Joseph~P Magliano.
\newblock Self-explanation and metacognition: The dynamics of reading.
\newblock In \emph{Handbook of metacognition in education}, pages 60--81. Routledge, 2009.

\bibitem[McNeil et~al.(2023)McNeil, Tran, A., J., Sarsa, Denny, Bernstein, and Leinonen]{mcneil23}
S.~McNeil, A.~Tran, Hellas A., Kim J., S.~Sarsa, P.~Denny, S.~Bernstein, and J.~Leinonen.
\newblock Experiences from using code explanations generated by large language models in a web software development e-book.
\newblock In \emph{Proceedings of the 54th ACM Technical Symposium on Computer Science Education V. 1.}, pages 931--937, Toronto, Ontario, Canada, 2023.

\bibitem[Mohler and Mihalcea(2009)]{mohler2009text}
Michael Mohler and Rada Mihalcea.
\newblock Text-to-text semantic similarity for automatic short answer grading.
\newblock In \emph{Proceedings of the 12th Conference of the European Chapter of the ACL (EACL 2009)}, pages 567--575, 2009.

\bibitem[Mohler et~al.(2011)Mohler, Bunescu, and Mihalcea]{mohler2011learning}
Michael Mohler, Razvan Bunescu, and Rada Mihalcea.
\newblock Learning to grade short answer questions using semantic similarity measures and dependency graph alignments.
\newblock In \emph{Proceedings of the 49th annual meeting of the association for computational linguistics: Human language technologies}, pages 752--762, 2011.

\bibitem[Oli et~al.(2023)Oli, Banjade, Lekshmi~Narayanan, Chapagain, Tamang, Brusilovsky, and Rus]{oli2023improving}
Priti Oli, Rabin Banjade, Arun~Balajiee Lekshmi~Narayanan, Jeevan Chapagain, Lasang~Jimba Tamang, Peter Brusilovsky, and Vasile Rus.
\newblock Improving code comprehension through scaffolded self-explanations.
\newblock In \emph{International Conference on Artificial Intelligence in Education}, pages 478--483. Springer, 2023.

\bibitem[OpenAI(2023)]{openai2023gpt4}
OpenAI.
\newblock Gpt-4 technical report.
\newblock \emph{CoRR}, abs/2303.08774, 2023.

\bibitem[Pontes et~al.(2018)Pontes, Huet, Linhares, and Torres-Moreno]{pontes2018predicting}
Elvys~Linhares Pontes, St{\'e}phane Huet, Andr{\'e}a~Carneiro Linhares, and Juan-Manuel Torres-Moreno.
\newblock Predicting the semantic textual similarity with siamese cnn and lstm.
\newblock \emph{arXiv preprint arXiv:1810.10641}, 2018.

\bibitem[Reimers and Gurevych(2019{\natexlab{a}})]{reimers-2019-sentence-bert}
Nils Reimers and Iryna Gurevych.
\newblock Sentence-bert: Sentence embeddings using siamese bert-networks.
\newblock In \emph{Proceedings of the 2019 Conference on Empirical Methods in Natural Language Processing}. Association for Computational Linguistics, 11 2019{\natexlab{a}}.
\newblock URL \url{http://arxiv.org/abs/1908.10084}.

\bibitem[Reimers and Gurevych(2019{\natexlab{b}})]{reimers2019sentence}
Nils Reimers and Iryna Gurevych.
\newblock Sentence-bert: Sentence embeddings using siamese bert-networks.
\newblock \emph{arXiv preprint arXiv:1908.10084}, 2019{\natexlab{b}}.

\bibitem[Sarsa et~al.(2022)Sarsa, Denny, Hellas, and Leinonen]{sarsa22}
S.~Sarsa, P.~Denny, A.~Hellas, and J.~Leinonen.
\newblock Automatic generation of programming exercises and code explanations using large language models.
\newblock In \emph{Proceedings of the 2022 ACM Conference on International Computing Education Research - Volume 1}, ICER '22, page 27–43, New York, NY, USA, 2022. Association for Computing Machinery.
\newblock ISBN 9781450391948.
\newblock \doi{10.1145/3501385.3543957}.
\newblock URL \url{https://doi.org/10.1145/3501385.3543957}.

\bibitem[Segura-Bedmar et~al.(2013)Segura-Bedmar, Mart{\'\i}nez~Fern{\'a}ndez, and Herrero~Zazo]{segura2013semeval}
Isabel Segura-Bedmar, Paloma Mart{\'\i}nez~Fern{\'a}ndez, and Mar{\'\i}a Herrero~Zazo.
\newblock Semeval-2013 task 9: Extraction of drug-drug interactions from biomedical texts (ddiextraction 2013).
\newblock Association for Computational Linguistics, 2013.

\bibitem[Sultan et~al.(2016)Sultan, Salazar, and Sumner]{sultan2016fast}
Md~Arafat Sultan, Cristobal Salazar, and Tamara Sumner.
\newblock Fast and easy short answer grading with high accuracy.
\newblock In \emph{Proceedings of the 2016 Conference of the North American Chapter of the Association for Computational Linguistics: Human Language Technologies}, pages 1070--1075, 2016.

\bibitem[Sung et~al.(2019)Sung, Dhamecha, Saha, Ma, Reddy, and Arora]{sung2019pre}
Chul Sung, Tejas Dhamecha, Swarnadeep Saha, Tengfei Ma, Vinay Reddy, and Rishi Arora.
\newblock Pre-training bert on domain resources for short answer grading.
\newblock In \emph{Proceedings of the 2019 Conference on Empirical Methods in Natural Language Processing and the 9th International Joint Conference on Natural Language Processing (EMNLP-IJCNLP)}, pages 6071--6075, 2019.

\bibitem[Tamang et~al.(2021)Tamang, Alshaikh, Khayi, Oli, and Rus]{tamang2021comparative}
Lasang~Jimba Tamang, Zeyad Alshaikh, Nisrine~Ait Khayi, Priti Oli, and Vasile Rus.
\newblock A comparative study of free self-explanations and socratic tutoring explanations for source code comprehension.
\newblock In \emph{Proceedings of the 52nd ACM Technical Symposium on Computer Science Education}, pages 219--225, 2021.

\bibitem[Touvron et~al.(2023)Touvron, Martin, Stone, Albert, Almahairi, Babaei, Bashlykov, Batra, Bhargava, Bhosale, et~al.]{llama2}
H.~Touvron, L.~Martin, K.~Stone, P.~Albert, A.~Almahairi, Y.~Babaei, N.~Bashlykov, S.~Batra, P.~Bhargava, S.~Bhosale, et~al.
\newblock Llama 2: Open foundation and fine-tuned chat models.
\newblock \emph{arXiv preprint arXiv:2307.09288}, 2023.

\bibitem[Wei et~al.(2022)Wei, Wang, Schuurmans, Bosma, Xia, Chi, Le, Zhou, et~al.]{wei2022chain}
Jason Wei, Xuezhi Wang, Dale Schuurmans, Maarten Bosma, Fei Xia, Ed~Chi, Quoc~V Le, Denny Zhou, et~al.
\newblock Chain-of-thought prompting elicits reasoning in large language models.
\newblock \emph{Advances in Neural Information Processing Systems}, 35:\penalty0 24824--24837, 2022.

\bibitem[Williams et~al.(2017)Williams, Nangia, and Bowman]{williams2017broad}
Adina Williams, Nikita Nangia, and Samuel~R Bowman.
\newblock A broad-coverage challenge corpus for sentence understanding through inference.
\newblock \emph{arXiv preprint arXiv:1704.05426}, 2017.

\bibitem[Zhang et~al.(2019)Zhang, Kishore, Wu, Weinberger, and Artzi]{zhang2019bertscore}
Tianyi Zhang, Varsha Kishore, Felix Wu, Kilian~Q Weinberger, and Yoav Artzi.
\newblock Bertscore: Evaluating text generation with bert.
\newblock \emph{arXiv preprint arXiv:1904.09675}, 2019.

\bibitem[Zhong et~al.(2023)Zhong, Ding, Liu, Du, and Tao]{zhong2023can}
Qihuang Zhong, Liang Ding, Juhua Liu, Bo~Du, and Dacheng Tao.
\newblock Can chatgpt understand too? a comparative study on chatgpt and fine-tuned bert.
\newblock \emph{arXiv preprint arXiv:2302.10198}, 2023.

\end{thebibliography}

\appendix

\section{Dataset Distribution} 
\label{D}

\begin{table}[H]
    \centering
    \begin{tabular}{c|c}
     Similarity Score(1-5)& No. of Sentence Pair   \\
    \hline
    1 & 529 (29.88\%) \\
    2 & 507 (28.62\%) \\
    3 & 419 (23.65\%) \\
    4 & 253 (14.45\%) \\
    5 & 62  (3.50\%) \\
    \end{tabular}
    \caption{Distribution of Data}
    \label{tab:data_distribution}
\end{table}

\section{Prompts}\label{apd:prompts}


\begin{tcolorbox}
    \textbf{Baseline Prompt [1-5]:} Analyze if the two sentences are similar and provide a score between 1 to 5,  with 1 indicating minimal similarity and 5 representing maximal similarity. Provide semantic similarity score for  \textit{<user explanation>} and \textit{<expert explanation>} between 1 to 5. Only provide the score without any other text.
\end{tcolorbox}

\begin{tcolorbox}
    \textbf{Baseline Prompt[0-1]:} Assess the similarity of the two sentences and assign a similarity score on a scale from 0 to 1, with 0 indicating minimal similarity and 1 representing maximal similarity. Provide semantic similarity score for \textit{<user explanation>} and \textit{<expert explanation>} between 0 to 1. Only provide the score without any other text. 
\end{tcolorbox}

\begin{tcolorbox}
    \textbf{Few Shot Prompt [0-1]:}
    Assess the similarity of the two sentences and assign a similarity score on a scale from 0 to 1, with 0 indicating minimal similarity and 1 representing maximal similarity. Provide a semantic similarity score for `Declares the array we want to use for our assignment' and `We initialize the array of type int to hold the specified numbers.' between 0 and 1. Only provide the score without any other text. Similarity Score: 0.87
    \newline

 Assess the similarity of the two sentences and assign a similarity score on a scale from 0 to 1, with 0 indicating minimal similarity and 1 representing maximal similarity. Provide semantic similarity score for `run a while-loop as long as the remainder of num/divisor is not equal to 0' 
   and `We could check whether the divisor is not a factor of the number by computing the remainder of the division of the number by the divisor.' between 0 and 1. Only provide the score without any other text. Similarity Score: 0.75
    \newline
 ............
 \newline
    Assess the similarity of the two sentences and assign a similarity score on a scale from 0 to 1, with 0 indicating minimal similarity and 1 representing maximal similarity.
    Provide semantic similarity score for \textit{<user explanation>} and \textit{<expert explanation>} between 0 to 1. Only provide the score without any other text. Similarity Score: 

\end{tcolorbox}

\begin{tcolorbox}
\textbf{Chain-of-Thought (CoT) Prompt [0-1]}: Discuss how these two texts are similar and different, then assign a semantic similarity score between [0.0-1.0] which describes their semantic similarity:
`Declares the array we want to use for our assignment, and `We initialize the array of type int to hold the specified numbers.
Similarity: Lets think step by step. Both the text is about declaration or initialization of array. The slight difference between the two texts is the second text provides additional information about the type of the declared array. Thus, these sentences have a [semantic similarity = 0.87]
\newline

Discuss how these two texts are similar and different, then assign a semantic similarity score between [0.0-1.0] which describes their semantic similarity:
`We could check whether the divisor is not a factor of the number by computing the remainder of the division of the number by the divisor.'	and 
`run a while-loop as long as the remainder of num/divisor is not equal to 0'
Similarity: Lets think step by step. Both the text is about computing the checking whether divisor is a factor of number or not.  However, the first text is more specific about using a while-loop and the condition for the loop to continue, while the second text is more focused on the purpose of the operation, which is to check if the divisor is a factor of the number. Thus, these sentences have a [semantic similarity = 0.75]
\newline

Discuss how these two texts are similar and different, then assign a semantic similarity score between [0.0-1.0] which describes their semantic similarity: `Loop start' and `We need to increment the divisor repeatedly as long as the divisor is not a factor of the number.Similarity: Lets think step by step. Both texts are discussing loop. The first text is simply stating the start of a loop, while the second text is explaining a specific condition within a loop. Thus, these sentences have a [semantic similarity = 0.2]
\newline

Discuss how these two texts are similar and different, then assign a semantic similarity score between [0.0-1.0] which describes their semantic similarity:\textit{<user explanation>} and \textit{<expert explanation>} 
Similarity: Lets think step by step

\end{tcolorbox}


\section{Chain-of-Though-Prompting Example}\label{apd:error_analysis}

\subsection{LLM response with semantic similarity and reasoning for C0T Prompting }
\label{apd:cot_response}

\begin{tcolorbox}
\textbf{Code statement:} System.out.println("The integer is positive.");	\\
\textbf{Expert explanation:} Print that the integer is positive if it is greater than 0.. \\
\textbf{Student explanation:} This statement prints that the integer is positive.\\
\textbf{Benchmark Similarity:} 1	 \\
\textbf{LLM:} GPT-4 \\
Both texts are discussing the same action of printing a statement that an integer is positive. The first text, however, provides a condition for this action (if the integer is greater than 0), while the second text does not specify any condition. Thus, these sentences have a [semantic similarity = 0.8]
have a [semantic similarity = 0.8]
\end{tcolorbox}


\begin{tcolorbox}
\textbf{Code statement:} int divisor = 2;	\\
\textbf{Expert explanation:} We define variable divisor to store the smallest divisor of the number. \\
\textbf{Student explanation:} we initialize another variable named divisor with the value 2.	\\
\textbf{Benchmark Similarity:} 0.33	 \\
\textbf{LLM:} GPT-4 \\
Both texts discuss initializing a variable named 'divisor', but they differ in the context and purpose of this variable. The first text simply states that the variable is initialized with the value 2, while the second text explains that the variable is used to store the smallest divisor of a number. Thus, these sentences have a [semantic similarity = 0.6]

\end{tcolorbox}

\subsection{Fail case: LLM fails in numerical reasoning}
\label{apd:error_analysis}

\begin{tcolorbox}
\textbf{Code statement:} int num = 15;	\\
\textbf{Expert explanation:} In this program, we initialize variable num to 15. \\
\textbf{Student explanation:} creates variable integer entitled "num" with initial value 5	\\
\textbf{Benchmark Similarity:} 0.466666667	 \\
\textbf{LLM:} GPT-4 \\
\textbf{LLM response to Chain-of-thought Prompting:} Both texts are discussing the initialization of a variable named "num", but the initial values they assign to "num" are different. The first text assigns a value of 5, while the second assigns a value of 15. Thus, these sentences have a [semantic similarity = 0.8] \\
\textbf{Comment:} LLM assigns high semantic equivalence score to the two text, which although linguistically similar involve different numerical value. In an educational context, this phenomenon may stem from potential misconceptions in students' understanding and may need targeted intervention to address the conceptual gap.

\end{tcolorbox}


\end{document}